\documentclass[%
 reprint,
 amsmath,amssymb,
 aps,
prl,dvipdfmx
]{revtex4-1}
\usepackage[dvipdfmx]{graphicx}
\usepackage{dcolumn}
\usepackage{bm}

\begin{document}

\preprint{APS/123-QED}

\title{Strong Anisotropic Spin-Orbit Interaction Induced in Graphene by Monolayer WS$_2$}

\author{T. Wakamura$^1$, F. Reale$^2$, P. Palczynski$^2$, S. Gu\'{e}ron$^1$, C. Mattevi$^2$}
\author{H. Bouchiat$^1$}%
 \email{helene.bouchiat@u-psud.fr}
\affiliation{$^1$Laboratoire de Physique des Solides, Universite Paris-Sud, 91400, Orsay, France}%

\affiliation{$^2$Department of Materials, Imperial College London, Exhibition Road, London, SW7 2AZ, United Kingdom}%


\date{\today}

\begin{abstract}
We demonstrate strong anisotropic spin-orbit interaction (SOI) in graphene induced by monolayer WS$_2$. Direct comparison between graphene/monolayer WS$_2$ and graphene/bulk WS$_2$ system in magnetotransport measurements reveals that monolayer transition metal dichalcogenide (TMD) can induce much stronger SOI than bulk. Detailed theoretical analysis of the weak-antilocalization curves gives an estimated spin-orbit energy ($E_{\rm so}$) higher than 10 meV. The symmetry of the induced SOI is also discussed, and the dominant $z$ $\rightarrow$ $-z$ symmetric SOI can only explain the experimental results. Spin relaxation by the Elliot-Yafet (EY) mechanism and anomalous resistance increase with temperature close to the Dirac point indicates Kane-Mele (KM) SOI induced in graphene.  
\end{abstract}

\pacs{Valid PACS appear here}
\maketitle


Spin-orbit interaction (SOI) is a crucial ingredient for designing new exotic electronic properties of quantum conductors. Depending on the crystalline symmetry SOI can have drastic effects on the band structure of a trivial conductor and transform it into a topological insulator (TI) \cite{hasan}. 
More than a decade ago Kane and Mele showed that graphene could be a model system for the formation of a 2D TI in the presence of on site (intrinsic) SOI \cite{kane1, kane2}. In this system SOI leads to the quantum spin Hall (QSH) state with the opening of a spin-orbit (SO) gap around the $K$ and $K'$ points and the formation of chiral spin polarized topological edge states. However, it was later determined that realistic SOI in graphene is too small (24 $\mu$eV \cite{gmitra1}) to realize this intriguing state. 

While many methods have been proposed theoretically and experimentally to enhance SOI of graphene such as hydrogenation \cite{castroneto, balakrishnan} and deposition of heavy elements \cite{weeks, calleja, brey, klimovskikh}, heterostructures with transition metal dichalcogenides (TMDs) are particularly interesting \cite{gmitra2, gmitra3, ulloa, kaloni, avsar2, wang, wang2, yang, yang2}. TMDs are two dimensional van der Waals materials similar to graphene but have intrinsic SOI of the order of 100 meV, much larger than that of graphene \cite{xiao}.  Monolayer TMDs have different band structures than those of bulk, and exhibit unique electrical, optical and mechanical properties \cite{klein, terrones, mak}. However, whereas the theoretical studies have focused on monolayer TMDs as a source for SOI, almost all experimental reports have studied multilayer TMDs and experiments on monolayer TMDs are lacking \cite{gmitra2, gmitra3, ulloa, kaloni, avsar2, wang, wang2, yang, yang2}. Moreover, direct comparison between monolayer and bulk TMDs for the efficient generation of strong SOI in graphene remains unexplored, and the nature of the induced SOI is still unclear \cite{gmitra2, gmitra3, ulloa, wang, wang2, yang, yang2}. \\
\begin{figure}[b]
\includegraphics[width=7.5cm,clip]{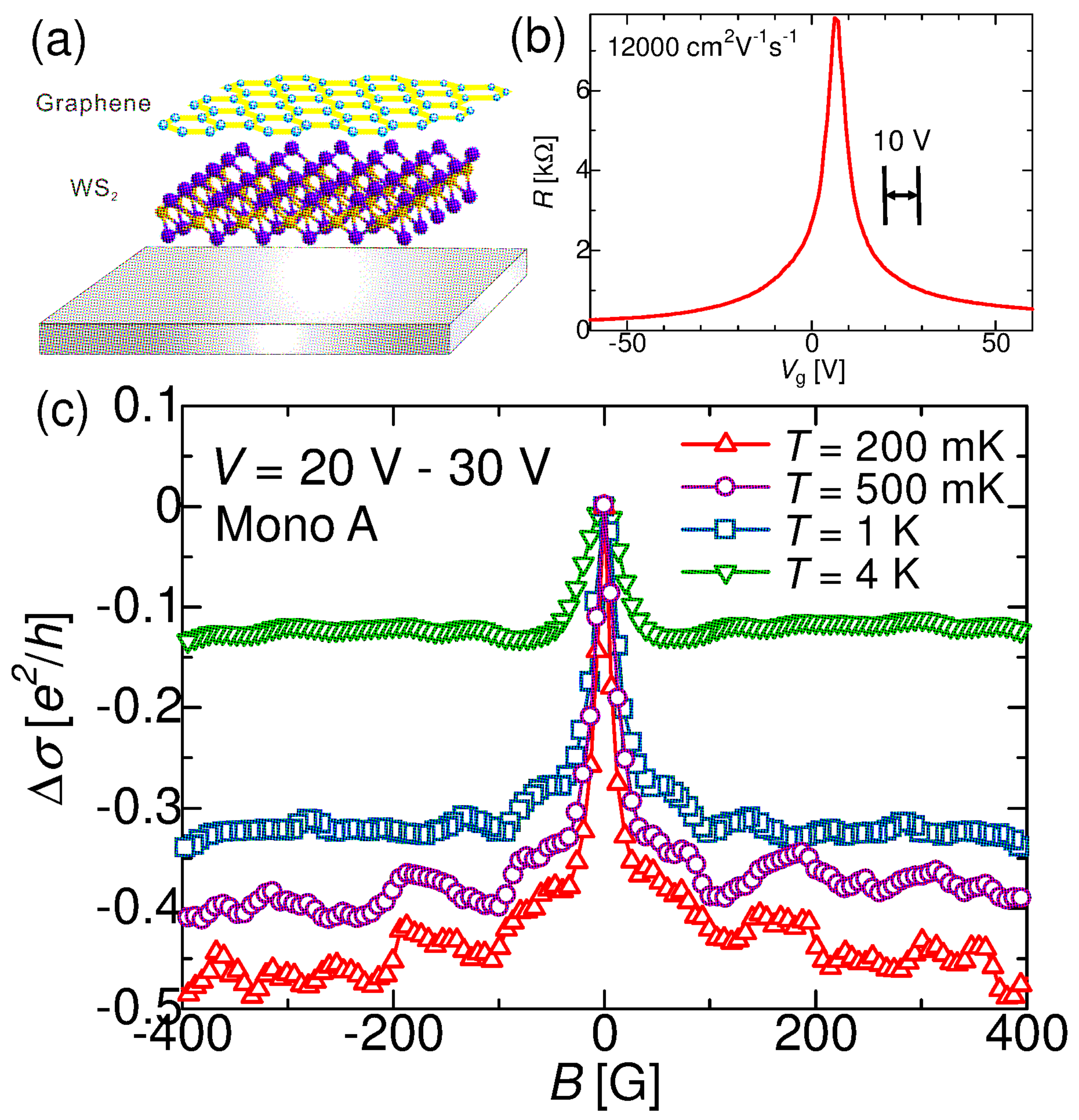}
\caption{Data taken from the Mono A. (a) Sketch of the G-mono samples. 
(b) Resistance as a function of the gate voltage. 
(c) $\Delta \sigma (B) \equiv \sigma (B) - \sigma (0)$ at three different temperatures. Without any subtraction the sharp WAL peak and flat tail are observed, signatures of the strong SOI induced in graphene. The $V_g$ range we average over for these experimental data is shown in (b) with the black marker.}
\label{fig1}
\end{figure}
\begin{figure*}[t]
\begin{center}
\includegraphics[width=15cm,clip]{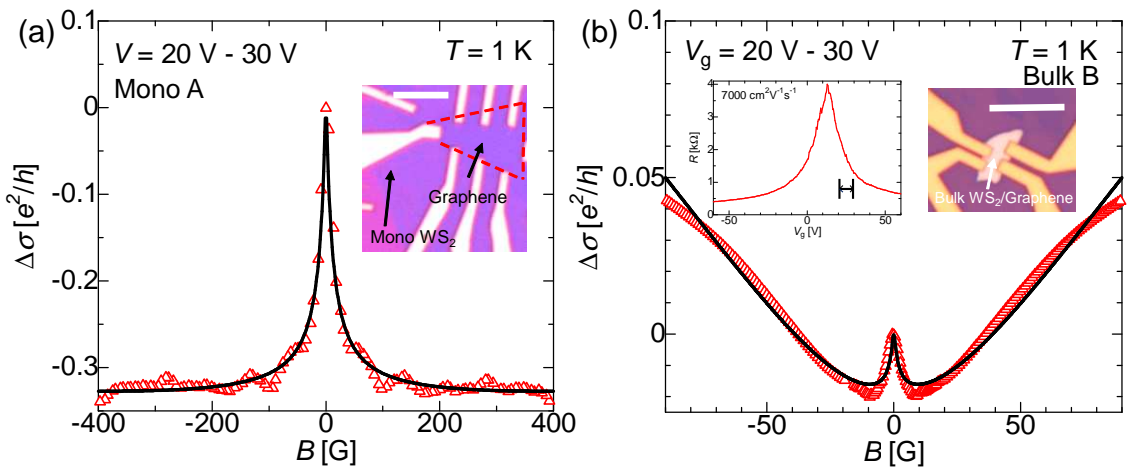}
\caption{Comparison of the magnetoconductivity curves for G-mono (a) and G-bulk (b) with the best theoretical fits (black solid curves). (a) $\Delta \sigma (B)$ from Mono A. The inset shows the optical image of the sample with the red dashed outline for graphene. (b) Magnetoconductivity curve for Bulk B averaged over a similar gate voltage range and similar doping level as that of (a) (see the left inset). 
Right inset shows the image of the Bulk B. The scale bar for the two sample images represents 10 $\mu$m.}
\label{Figure2}
\end{center}
\end{figure*}
In this Letter, we demonstrate a striking difference in the capacity of monolayer and bulk WS$_2$ to induce SOI in graphene: Monolayer WS$_2$ can induce much stronger SOI in graphene than bulk WS$_2$. At low temperatures magnetotransport measurements display clear weak-antilocalization (WAL) peaks for both graphene/monolayer WS$_2$ (G-mono) and graphene/bulk WS$_2$ (G-bulk) systems, regardless of the carrier type. For the G-mono system, the magnetoconductance curves display a large peak at low magnetic fields and remain remarkably flat at high magnetic fields, and the estimated SOI by theoretical analysis \cite{mccann} is greater than 10 meV. This value is an order of magnitude larger than SOI induced in the G-bulk system. We elucidate the symmetry of induced SOI and find that $z \rightarrow -z$ symmetric SOI is much stronger than the asymmetric one. 
Detailed analysis on spin relaxation mechanism shows a large contribution from Kane-Mele (KM) SOI to spin relaxation close to the Dirac point. An anomalous temperature increase of the resistance between room temperatures and 77 K suggests an induced spin-orbit gap. These evidences indicate that not only valley-Zeeman (VZ) but also large Kane-Mele (KM) type SOI is induced in graphene.

Two different types of heterostructures are prepared for this study: monolayer WS$_2$/graphene and bulk WS$_2$/graphene. Graphene is mechanically exfoliated from natural graphite. Monolayer WS$_2$ flakes are grown by chemical vapour deposition (CVD) directly on a silicon substrate, and then transferred onto another Si/SiO$_2$ substrate to avoid defects induced in the SiO$_2$ layer. Graphene is picked up by polymethyl methacrylate (PMMA) then deposited onto WS$_2$. Bulk WS$_2$ is prepared by mechanical exfoliation and deposited on graphene. Conventional electron beam lithography techniques are employed to form electrical contacts (Ti (5 nm)/Au (100 nm)). Several samples are fabricated for both types, and in this Letter we focus on two samples for the G-mono type (Mono A and B) and G-bulk type (Bulk A and B). We note that for all samples the whole area of graphene is on WS$_2$. Details of the fabrication are given in the Supplemental Material \cite{suppl}.

We first show the experimental results of Mono A. The structure of the samples is schematically illustrated in Fig. 1(a). Figure 1(b) shows the gate voltage ($V_g$) dependence of resistance ($R$). 
We note that the resistivity of the WS$_2$ underlayer is much larger than that of graphene and the charge transport is mostly dominated by graphene at low temperatures. 
The mobility of the present sample is 12000 cm$^2$V$^{-1}$s$^{-1}$ at 200 mK.

For evaluating the induced SOI in graphene, we employ magnetotransport measurements with magnetic fields perpendicular to the graphene plane. 
Observation of the regime (weak localization (WL) or weak antilocalization (WAL)) in magnetotransport measurements reveals the amplitude of SOI \cite{wang, wang2, yang, yang2}.
Figure 1(c) shows the quantum conductivity correction ($\Delta \sigma (B) \equiv \sigma (B) - \sigma (0)$) as a function of a magnetic field for the electon-doped region at different temperatures. To suppress the effect of universal conductance fluctuation (UCF), we average 50 curves with different $V_g$ in a 10 V range around a given $V_g$. The sample exhibits a clear WAL peak, beyond which the magnetoconductivity curve is extremely flat even for higher field regions. The gradient of $\Delta \sigma (B)$ in the high field region is determined by the competition between WL and WAL effect. When $\Delta \sigma (B)$ increases with $B$ in this region, the WL is dominant while with increasing WAL the gradient of $\Delta \sigma (B)$ becomes small, and flat in the case of strong SOI \cite{bergmann}. Therefore, the flat $\Delta \sigma (B)$ curve in Fig. 1(c) for large $B$ already reveals the strong SOI induced in graphene by the monolayer WS$_2$. 
We find similar shapes with flat tails in a high field region for the other gate voltage ranges in the electron-doped regime. 
Our results are without any subtractions of the background signals carried out in \cite{wang2} to suppress the gradient of $\Delta \sigma (B)$.  
To estimate the SOI amplitude, we fit the data using the theoretical expression given in \cite{mccann}. We note that in the temperature range that we explore, the WAL driven by the pseudospin-orbit coupling is suppressed since the phase coherence time is much longer than the intervalley scattering time \cite{Tihomenko}. As pointed out in other studies \cite{wang2, grbic}, in the case with the flat tails in $\Delta \sigma (B)$ for large $B$ it is essential to take into account not only the small magnetic field region but also the higher region to determine the spin-orbit time ($\tau_{\rm so}$) accurately. The theoretical expression is \cite{mccann}
\begin{multline}
\Delta \sigma (B) = -\frac{e^2}{2 \pi h} \left[F \left( \frac{\tau_B^{-1}}{\tau_\phi^{-1}} \right) - F \left( \frac{\tau_B^{-1}}{\tau_\phi^{-1} + 2 \tau_{\rm asy}^{-1}} \right) \right.\\
\left. -2 F \left( \frac{\tau_B^{-1}}{\tau_\phi^{-1} + \tau_{\rm so}^{-1}} \right) \right],
\end{multline}
where $F(x) = \ln(x)+\psi(1/2 + 1/x)$, with $\psi(x)$ the digamma function. $\tau_{\rm so}^{-1} = \tau_{\rm sym}^{-1} + \tau_{\rm asy}^{-1}$, where sym (asy) denotes the symmetric (asymmetric) contribution to the SOI (discussed below in detail). The fits yield three parameters $\tau_\phi$, $\tau_{\rm asy}$ and $\tau_{\rm so}$. $\tau_{\rm so}$ determines the total amplitude of SOI in the system, and from $\tau_{\rm asy}$ one can evaluate the symmetry type of SOI.

\begin{figure}[t]
\includegraphics[width=7.5cm,clip]{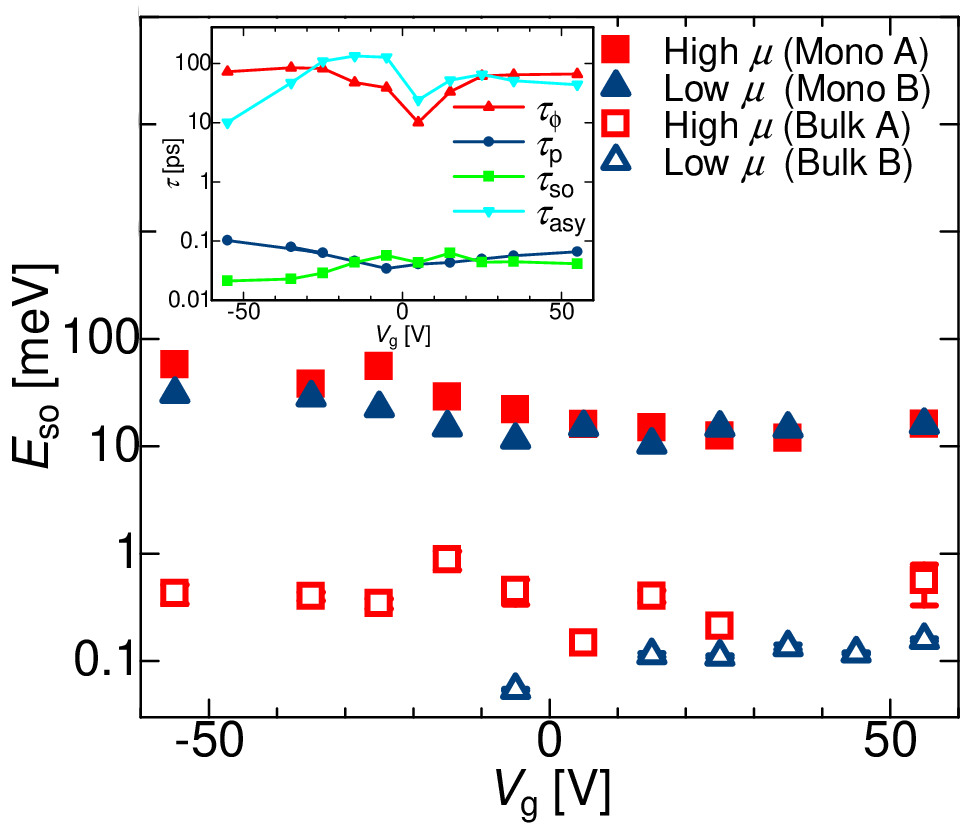}
\caption{Comparison of the spin-orbit energy ($E_{\rm so}$) estimated from the theoretical fitting of both G-mono and G-bulk. For each type, the data taken from the two samples with different mobilities ($\mu$) are shown. The mobilities are 12000 cm$^2$V$^{-1}$s$^{-1}$ (Mono A), 7000 cm$^2$V$^{-1}$s$^{-1}$ (Mono B), 9000 cm$^2$V$^{-1}$s$^{-1}$ (Bulk A) and 7000 cm$^2$V$^{-1}$s$^{-1}$ (Bulk B). 
Inset: Fitting parameters obtained for the Mono B. $\tau_{\rm asy}$ is found to be comparable to $\tau_\phi$ and much larger than $\tau_{\rm so}$.}
\label{fig1}
\end{figure}

We here focus on the amplitude of the induced total SOI ($\tau_{\rm so}$) and discuss the nature of SOI in the later section. As shown in Fig. 2(a), the fitting reproduces well the experimental data when $\tau_{\rm so} = 0.05$ ps. This corresponds to the spin-orbit energy ($E_{\rm so} \equiv \hbar / \tau_{\rm so}$) equal to 13 meV. This value is much larger than those reported in previous studies \cite{wang, yang, yang2}. We note that this value is smaller than that reported in \cite{avsar2}, but they assume the EY spin relaxation mechanism and use different definition for spin-orbit energy.

We next turn to the magnetoconductivity of the G-bulk samples to compare the efficiency for inducing stronger SOI in graphene. 
The experimental data of $\Delta \sigma (B)$ as a function of $B$ exhibit clear WAL peaks as observed for the G-mono samples (Fig. 2(b)). However, in contrast to the G-mono samples, even for small magnetic fields the magnetoconductivity curve shows a steep upturn. The theoretical fitting based on (1) yields $\tau_{\rm so}$ = 4 ps as a best fit, consistent with the previous reports \cite{avsar2, wang}. This value corresponds to $E_{\rm so}$ = 170 $\mu$eV, much smaller than the estimated value of $E_{\rm so}$ for the G-mono samples. In Fig. 3, we compare $E_{\rm so}$ as a function of $V_g$ for the two G-mono and G-bulk samples with different mobilities. Strikingly, there is an order of magnitude difference in the amplitude of the induced SOI in graphene between the G-mono and G-bulk samples, and regardless of the mobilities the G-mono samples show extremely strong SOI for both electron-doped and hole-doped region. 
We note that no background signals are subtracted from the original data for the electron-doped region of the sample Mono A and two bulk samples \cite{suppl}. These results explicitly demonstrate that monolayers of TMDs can more efficiently induce strong SOI in graphene than the bulk TMDs.

It is important to elucidate not only the amplitudes but also the nature of the induced SOI. As studied in \cite{mccann}, there are two types of SOI with different symmetry in real space which can contribute in our systems. One is symmetric in the $z \rightarrow -z$ inversion, where the $z$ axis is normal to the graphene plane. The other contribution is asymmetric in the $z \rightarrow -z$ inversion.
As expressed in (1), from the theoretical fitting the two different parameters ($\tau_{\rm sym}$ and $\tau_{\rm asy}$) relevant to each contribution can be obtained. In Fig. 4, we show the theoretical fits with the different $E_{\rm so}$ for the Mono A sample. For the best fit $E_{\rm so}$ = 13 meV and $E_{\rm asy}$ = 0.059 meV, indicating that the symmetric SOI strongly dominates the induced SOI in graphene. In the inset of Fig. 4(a), the three terms in the equation (1) are visualized for the case of the best fit. This plot reveals that the first and second terms play a dominant role for the fitting. To reproduce the experimentally observed sharp peak around $B$ = 0 and flat $\Delta \sigma (B)$ for the higher $B$ region, we find that a $\tau_{\rm asy}$ comparable to $\tau_\phi$ and much smaller $\tau_{\rm so}$ are essential. As shown in Fig. 4(a), increasing $\tau_{\rm so}$ and decreasing $\tau_{\rm asy}$ gives rise to the upturn of $\Delta \sigma (B)$ for larger $B$. Due to the negligible contribution of the third term when $\tau_{\rm so}$ is extremely small, our estimation of $\tau_{\rm so}$ only gives an upper bound, and the actual SOI could be even larger. 

We also investigate spin relaxation mechanisms of our systems. In graphene two spin relaxation mechanisms are possible, the EY and D'yakonov-Perel (DP) mechanism. The two mechanisms can be identified by different dependence of $\tau_{\rm so}$ on $\tau_p$: For the EY (DP) mechanism $\tau_{\rm so} \propto \tau_p$ $(\tau_p^{-1})$. 
As demonstrated in the previous study \cite{zomer}, by fitting the relation between $\tau_p$ and $\tau_{\rm so}$ following the equation
\begin{equation}
\frac{\varepsilon_F^2 \tau_p}{\tau_{\rm so}} = \Delta_{\rm EY}^2 + \left( \frac{4 \Delta_{\rm DP}^2}{\hbar^2} \right)
\varepsilon_F^2 \tau_p^2
\label{eq1}
\end{equation}
where $\Delta_{\rm EY(DP)}$ is the amplitude of spin-orbit coupling due to the EY (DP) mechanism and $\varepsilon_F$ is the Fermi energy, one can indentify the dominant spin relaxation mechanism. Figure 4(b) shows the relation between $\varepsilon_F^2 \tau_p/\tau_{\rm so}$ and $\varepsilon_F^2 \tau_p^2$ using the experimental results and the theoretical fits. The fits for G-mono samples give $\Delta_{\rm EY}$ = 20 - 27 meV while $\Delta_{\rm DP}$ = 4 - 6 meV. Since $\Delta_{\rm EY}$ is much larger than $\Delta_{\rm DP}$, the EY mechanism is dominant for spin relaxation in our system particularly close to the Dirac point, where $\tau_p$ is small. Interestingly, the obtained $\Delta_{\rm EY}$ is in the same order of magnitude as the value of $E_{\rm so}$ given by analysis of WAL. On the other hand, for G-bulk samples the DP contribution is one order of magnitude smaller than those for G-mono samples ($\Delta_{\rm DP} \sim$ 0.7 meV) and the EY contribution is so small that the extrapolation of (2) even takes a small negative value. 

The identification of the dominant spin relaxation mechanism provides important information on the nature of the induced SOI. 
The original theory by Kane and Mele requires dominant $z \rightarrow -z$ symmetric SOI \cite{kane2} for the QSHE to be realized, and recent theories predict that in the case of graphene on TMD, symmetric SOI is further classified as two different contributions by different symmetry in sublattice space; KM and valley-Zeeman (VZ) SOI \cite{gmitra3, roche, valenzuera, vanwees, ulloa, frank}. The KM term is proportional to the $z$ component of the pseudospin ($\sigma_z$) whereas VZ SOI depends on the unit matrix in sublattice space ($\sigma_0$). VZ SOI is driven by broken sublattice symmetry in graphene, and provides the DP spin relaxation mechanism \cite{roche}. In contrast, symmetric nature of the KM SOI yields EY spin relaxation \cite{mccann}. The large EY spin relaxation close to the Dirac point in our system thus indicates that the induced SOI has large contribution from the KM-type SOI. We note that the DP mechanism includes two contributions, VZ and Rashba SOI, and yields anisotropic spin relaxation in highly doped region \cite{roche}. For detailed discussions please see \cite{suppl}.
\begin{figure}[t]
\includegraphics[width=8.5cm,clip]{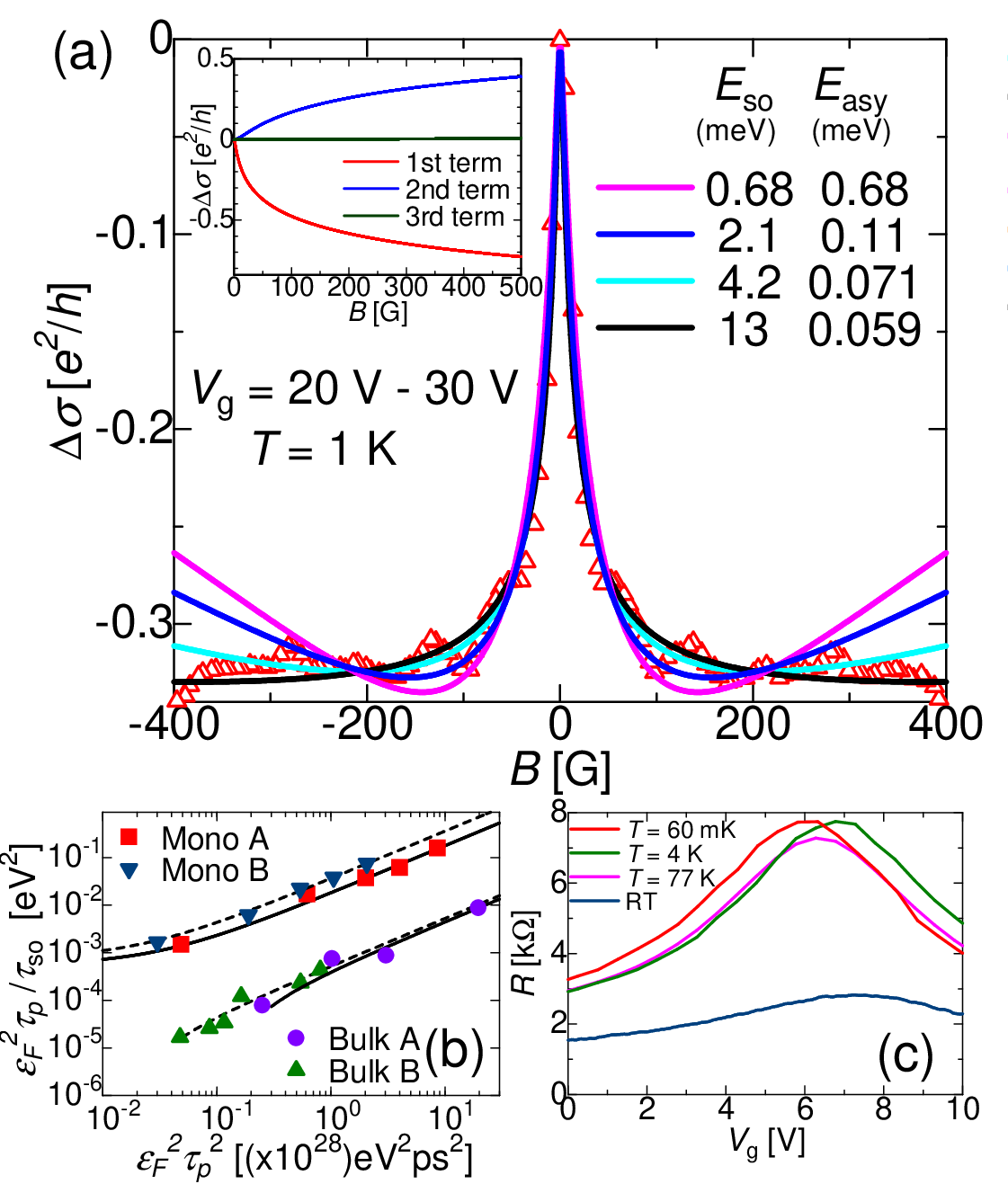}
\caption{(a) The experimental results and fitting curves with different $E_{\rm so}$ and $E_{\rm asy}$ for Mono A. Larger $E_{\rm so}$ and smaller $E_{\rm asy}$ give a better fit. Inset: Three different terms in the equation (1) as a function of $B$ for the best fit. (b) $\varepsilon_F^2 \tau_p/\tau_{\rm so}$ as a function of $\varepsilon_F^2 \tau_p^2$ for G-mono and G-bulk. Theoretical fits based on (2) is shown with the solid(dashed) line for Mono and Bulk A(B). (c) Gate voltage dependence of registance at different temperatures taken from Mono A.}
\label{fig1}
\end{figure}

To obtain other signatures of induced SOI, we measure the temperature dependence of the resistance ($R$) of graphene for both G-mono and G-bulk samples between 60 mK and room temperature (RT). Surprisingly, for the G-mono samples we observe a strong increase of $R$ with decreasing temperature around the Dirac point. Especially for Mono A, $R$ increases by a factor of three close to the Dirac point between RT and 77 K, and saturates below 77 K (Fig. 4(c)). In contrast, the G-bulk system's resistance varies by only about 30 \% even between RT and 60 mK. Pristine graphene with mobility comparable to our samples has been reported to exhibit a weak temperature dependence of resistance \cite{morozov, tan}. Therefore, the anomalous increase of $R$ with decreasing temperature suggests a gap-opening around the Dirac point for the bulk states and strong interaction between monolayer WS$_2$ and graphene compared with that from bulk. Interestingly, the fitting of the temperature increase between RT and 77 K based on a simple semiconductor model gives the energy gap $E_g \sim 16$ meV, comparable to the estimated $E_{\rm so}$ values. Saturation below 77 K indicates the existence of a small number of residual conducting states in the gap.
Detailed discussions are given in the Supplemental Material \cite{suppl}

Based on these experimental results, large symmetric SOI induced in G-mono samples may include large KM-type component similar to the one proposed by Weeks \cite{weeks}. Theoretically, graphene on TMD has spin-orbit potential ($\lambda_{\rm I}^\alpha$, $\alpha$= A or B) induced by TMD different at sublattice A and B ($\lambda_{\rm I}^{\rm A} \neq \lambda_{\rm I}^{\rm B}$) \cite{gmitra2, gmitra3, frank}. In contrast, experimentally, due to the large incommensurability of the lattice constants of graphene and WS$_2$, electrons (or holes) at sublattice A and B feel similar spin-orbit potential averaged over space. This effect is stronger in low doped region (close to the Dirac point) because the Fermi wave length becomes longer. The averaged spin-orbit potential is preferable to enhance KM SOI rather than VZ SOI, consistent with the dominant EY spin relaxation mechanism close to the Dirac point and large temperature increase. Stronger SOI for G-mono samples than G-bulk ones can be due to different band structures. 
Since graphene's electronic transport is dominated by carriers around the $K$ (or $K'$) point, monolayer TMDs, which have a direct band gap at $K$ and $K'$ points, may have stronger interaction with graphene. 

In conclusion, we successfully induced strong SOI more than 10 meV in graphene with monolayer WS$_2$. Direct comparison with the bulk WS$_2$ system reveals higher efficiency of monolayer WS$_2$ to induce much stronger SOI in graphene. The SOI is dominantly of the symmetric type, and analysis on spin relaxation mechanism demonstrates existence of KM SOI which dominates spin relaxation around the Dirac point. The strong increase of the resistance with decreasing temperature for the graphene/monolayer WS$_2$ samples also supports existence of KM SOI. 

We gratefully acknowledge very useful discussions with A. Zobelli, M. Goerbig, J. Meyer, S. Ilic, A. W. Cummings, J. H. Garcia, S. Roche and R. Deblock, and help for the sample fabrication by S. Sengupta and A. Murani. This project is financially supported in part by the Marie Sklodowska Curie Individual Fellowships, the ANR grants DIRACFORMAG, MAGMA, JETS, the CNRS and the award of a Royal Society University Research Fellowship by the UK Royal Society, the EPSRC grant EP/M022250/1 and the EPSRC-Royal Society Fellowship Engagement Grant EP/L003481/1.





\begin{thebibliography}{40}



\bibitem{hasan} M. Z. Hasan and C. L. Kane, Rev. Mod. Phys. \textbf{82}, 3045 (2010).

\bibitem{kane1} C. L. Kane and E. J. Mele, Phys. Rev. Lett. \textbf{95}, 226801 (2005).

\bibitem{kane2} C. L. Kane and E. J. Mele, Phys. Rev. Lett. \textbf{95}, 146802 (2005).

\bibitem{gmitra1} M. Gmitra, S. Konschuh, C. Ertler, C. Ambrosch-Draxl and J. Fabian, Phys. Rev. B \textbf{80}, 235431 (2009).

\bibitem{castroneto} A. H. Castro Neto and F. Guinea, Phys. Rev. Lett. \textbf{103}, 026804 (2009).

\bibitem{balakrishnan} J. Balakrishnan, G. K. W. Koon, M. Jaiswal, A. H. Castro Neto and B. \"{O}zyilmaz, Nat. Phys. \textbf{9}, 284 (2013).

\bibitem{weeks} C. Weeks, J. Hu, J. Alicea, M. Franz and R. Wu, Phys. Rev. X \textbf{1}, 021001 (2011).

\bibitem{calleja} F. Calleja, \textit{et al}., Nat. Phys. \textbf{11}, 43 (2015).

\bibitem{brey} L. Brey, Phys. Rev. B \textbf{92}, 235444 (2015).

\bibitem{klimovskikh} I. I. Klimovskikh \textit{et al}., ACS Nano \textbf{11}(1), 368 (2017).

\bibitem{gmitra2} M. Gmitra and J. Fabian, Phys. Rev. B \textbf{92}, 155403 (2015).

\bibitem{gmitra3} M. Gmitra, D. Kochan, P. H\"{o}gl, and J. Fabian, Phys. Rev. B \textbf{93}, 155104 (2016).

\bibitem{ulloa} A. M. Alsharari, M. M. Asmar, and S. E. Ulloa, Phys. Rev. B \textbf{94}, 241106(R) (2016).

\bibitem{kaloni} T. P. Kaloni, L. Kou, T. Frauenhaim and U. Schwingenschl\"{o}gl, Appl. Phys. Lett. \textbf{105}, 233112  (2014).

\bibitem{avsar2} A. Avsar \textit{et al}., Nat. Commun. \textbf{5}, 4875 (2014).

\bibitem{wang} Z. Wang, D. K. Ki, H. Chen, H. Berger, A. H. MacDonald  and  A. F. Morpurgo, Nat. Commun. \textbf{6}, 8339 (2015).

\bibitem{wang2} Z. Wang, D. K. Ki, J. Y. Khoo, D. Mauro, H. Berger, L. S. Levitov, and A. F. Morpurgo, Phys. Rev. X \textbf{6}, 041020 (2016).

\bibitem{yang} B. Yang \textit{et al}., 2D Mater. \textbf{3}, 031012 (2016).

\bibitem{yang2} B. Yang, M. Lohmann, D. Barroso, I. Liao, Z. Lin, Y. Liu, L. Bartels, K. Watanabe, T. Taniguchi and J. Shi, Phys. Rev. B \textbf{96}, 041409 (R) (2017).

\bibitem{xiao} D. Xiao, G. B. Liu, W. Feng, X. Xu and W. Yao, Phys. Rev. Lett. \textbf{108}, 196802 (2012).

\bibitem{klein} A. Klein, S. Tiefenbacher, V. Eyert, C. Pettenkofer and W. Jaegermann, Phys. Rev. B \textbf{64}, 205416 (2001).


\bibitem{terrones} H. Terrones, F. Lo\'{o}pez-Urias and M. Terrones, Sci. Rep. \textbf{3}, 1549 (2013).

\bibitem{mak} K. F. Mak, C. Lee, J. Hone, J. Shan and T. F. Heinz, Phys. Rev. Lett. \textbf{105}, 136805 (2010).
\bibitem{mccann} E. McCann and V. I. Fal'ko, Phys. Rev. Lett. \textbf{108}, 166606 (2012).

\bibitem{suppl} See Supplemental Material [URL] for more details of the sample fabrication, additional experimental data and their analysis, which includes \cite{altshuler},\cite{tikhonenko2}.

\bibitem{altshuler} B. L. Altshuler, A. G. Aronov, and D. E. Khmelnitsky, J. Phys. C \textbf{15}, 7367 (1982).

\bibitem{tikhonenko2} F. V. Tikhonenko, D. W. Horsell, R. V. Gorbachev, and A. K. Savchenko, Phys. Rev. Lett. \textbf{100}, 056802 (2008).

\bibitem{bergmann} G. Bergmann, Phys. Rep. \textbf{107}, 1 (1984).

\bibitem{Tihomenko} F. V. Tikhonenko, A. A. Kozikov, A. K. Savchenko and R. V. Gorbachev, Phys. Rev. Lett. \textbf{103}, 226801 (2009).

\bibitem{grbic} B. Grbi\'{c}, R. Leturcq, T. Ihn, K. Ensslin, D. Reuter and A. D. Wieck, Phys. Rev. B \textbf{77}, 125312 (2008).

\bibitem{zomer} P. J. Zomer, M. H. D. Guimar\~{a}es, N. Tombros and B. J. van Wees, Phys. Rev. B \textbf{86}, 161416(R) (2012).



\bibitem{roche} A. W. Cummings, J. H. Garcia, J. Fabian, and S. Roche, Phys. Rev. Lett. \textbf{119}, 206601 (2017).

\bibitem{valenzuera} L. A. Benitez, J. F. Sierra, W. S. Torres, A. Arrighi, F. Bonnell, M. V. Costache and S. O. Valenzuera, Nat. Phys. doi:10.1038/s41567-017-0019-2 (2017).

\bibitem{vanwees} T. S. Ghiasi, J. Ingla-Ayn\'{e}s, A. A. Kaverzin and B. J. van Wees. Nano Lett. \textbf{17}, 7528 (2017).

\bibitem{frank} T. Frank, P. H\"{o}gl, M. Gmitra, D. Kochan and J. Fabian, arXiv:1707.02124.




\bibitem{morozov} S. V. Morozov, K. S. Novoselov, M. I. Katsnelson, F. Schedin, D. Elias, J. A. Jaszczak and A. K. Geim, Phys. Rev. Lett. \textbf{100}, 016602 (2008).

\bibitem{tan} Y. -W. Tan, Y. Zhang, H. L. Stormer and P. Kim, Eur. Phys. J. Special Topics \textbf{148}, 15 (2007).







\end{thebibliography}
\end{document}